\documentclass[paper]{JHEP3}
\pdfoutput=1
\usepackage{amsmath,amssymb,amsthm,amscd,graphicx}
\input epsf.sty

\newtheorem{theorem}{Theorem}[section]

\theoremstyle{definition}

\newtheorem{remark}[theorem]{Remark}

\newcommand{\CC}{{\cal C}}

\newcommand{\CF}{{\cal F}}

\newcommand{\CN}{{\cal N}}
\newcommand{\CO}{{\cal O}}
\newcommand{\CP}{{\cal P}}

\newcommand{\CR}{{\cal R}}

\def\IZ{{\mathbb Z}}

\def\IP{{\mathbb P}}

\def\IS{{\mathbb S}}
\def\IF{{\mathbb F}}

\newcommand{\tr}{{\rm Tr}}
\newcommand{\re}{{\rm e}}
\newcommand{\ri}{{\rm i}}
\newcommand{\rd}{{\rm d}}
\renewcommand{\d}{\partial}

\newcommand{\be}{\begin{equation}}
\newcommand{\ee}{\end{equation}}
\newcommand{\ba}{\begin{aligned}}
\newcommand{\ea}{\end{aligned}}
\newcommand{\ben}{\begin{eqnarray}\displaystyle}
\newcommand{\een}{\end{eqnarray}}

\newcommand{\sectiono}[1]{\section{#1}\setcounter{equation}{0}}

\newdimen\tableauside\tableauside=1.0ex
\newdimen\tableaurule\tableaurule=0.4pt
\newdimen\tableaustep
\def\phantomhrule#1{\hbox{\vbox to0pt{\hrule height\tableaurule width#1\vss}}}
\def\phantomvrule#1{\vbox{\hbox to0pt{\vrule width\tableaurule height#1\hss}}}
\def\sqr{\vbox{%
  \phantomhrule\tableaustep
  \hbox{\phantomvrule\tableaustep\kern\tableaustep\phantomvrule\tableaustep}%
  \hbox{\vbox{\phantomhrule\tableauside}\kern-\tableaurule}}}
\def\squares#1{\hbox{\count0=#1\noindent\loop\sqr
  \advance\count0 by-1 \ifnum\count0>0\repeat}}
\def\tableau#1{\vcenter{\offinterlineskip
  \tableaustep=\tableauside\advance\tableaustep by-\tableaurule
  \kern\normallineskip\hbox
    {\kern\normallineskip\vbox
      {\gettableau#1 0 }%
     \kern\normallineskip\kern\tableaurule}%
  \kern\normallineskip\kern\tableaurule}}
\def\gettableau#1{\ifnum#1=0\let\next=\null\else
\squares{#1}\let\next=\gettableau\fi\next}

\newcommand{\figref}[1]{Fig.~\protect\ref{#1}}
\title{Exact results in ABJM theory from topological strings}

\author{
Marcos Mari\~no$^{a,b}$ and Pavel Putrov$^b$
\\
$^a$D\'epartement de Physique Th\'eorique et $^b$Section de Math\'ematiques,\\
Universit\'e de Gen\`eve, Gen\`eve, CH-1211 Switzerland\\
\\
\email{marcos.marino@unige.ch}, \quad
\email{pavel.putrov@unige.ch}
}

\abstract{Recently, Kapustin, Willett and Yaakov have found, by using localization techniques, 
that vacuum expectation values of Wilson loops in ABJM theory can be calculated with a matrix model. 
We show that this matrix model is closely related to Chern--Simons theory on a lens space with a gauge 
supergroup. This theory has a topological string large $N$ 
dual, and this makes possible to solve the matrix model exactly in the large $N$ expansion. In particular, we find the 
exact expression for the vacuum expectation 
value of a 1/6 BPS Wilson loop in the ABJM theory, as a function of the 't Hooft parameters, and in the planar limit. This expression 
gives an exact interpolating function between the weak and the strong coupling regimes. The behavior at strong coupling is in precise 
agreement with the prediction of the AdS string dual. We also give explicit results for the 1/2 BPS Wilson loop recently constructed by Drukker and Trancanelli.}    

\begin{document}

\sectiono{Introduction}
Localization techniques in supersymmetric gauge theories have produced in recent years explicit expressions for a variety of correlation 
functions. In \cite{pestun}, they were used to prove the longstanding conjecture \cite{dg,esz} that the vev of a 1/2 BPS Wilson loop 
in $\CN=4$ super Yang--Mills theory can be computed by a correlator in a Gaussian matrix integral. This gives the celebrated formula 
\be
\label{n4}
\langle W_{\tableau{1}}\rangle={2\over {\sqrt{\lambda}}} I_1\left({\sqrt{\lambda}}\right) 
\ee
for the planar limit of the Wilson loop in the fundamental representation. In this formula, $I_1$ is a modified Bessel function and $\lambda=g^2N$ is the 't Hooft coupling. (\ref{n4}) gives an exact expression in $\lambda$ that interpolates between strong and 
weak coupling (see \cite{sz} for a review). 

More recently, Kapustin, Willett and Yaakov \cite{kapustin} have applied localization techniques to ABJM theories \cite{abjm,abj}. These are superconformal 
field theories in three dimensions based on a $U(N_1)\times U(N_2)$ Chern--Simons theory coupled to matter, and they 
have large $N$ AdS$_4$ duals. In \cite{kapustin}, it was shown that the 
calculation of vevs of 1/6 BPS Wilson loops in these theories can be reduced to a calculation in a matrix model, 
and they verified that the Wilson loop vev reproduces the calculations 
of \cite{dp,cw,rey}. However, the matrix model they obtained is 
highly nontrivial, and the comparison with the gauge theory calculation was done by performing a perturbative expansion up 
to two loops. Some partial results on the planar limit of the matrix model have been obtained in \cite{suyama}.

In this note we point out that the matrix model of \cite{kapustin} can be solved by relating it to the 
Chern--Simons matrix models introduced in \cite{mm}, and in particular 
to the lens space matrix model studied in detail in \cite{akmv,hy}. In \cite{akmv} it was shown that the lens space 
Chern--Simons matrix model is the large $N$ dual of 
topological string theory on a certain class of local Calabi--Yau geometries, 
providing in this way a nontrivial generalization of the Gopakumar--Vafa duality \cite{gv}. 
This means in particular that the large $N$ limit of the model can be studied by using standard techniques in mirror 
symmetry. It turns out that the matrix model of \cite{kapustin} can be regarded as a 
supergroup version of the lens space matrix model studied in \cite{mm,akmv,hy}. 
Since large $N$ duals describe matrix models as well as their supergroup extensions, we can use topological string theory on a local Calabi--Yau 
geometry to obtain exact results in ABJM theory. 

In this paper we use the solution of \cite{akmv,hy} to obtain an {\it exact} expression 
for the planar limit of the vev of the 1/6 BPS Wilson loop constructed in \cite{dp,rey,cw}, as a function of the two 't Hooft parameters of the theory. 
This analytic expression is relatively complicated but it can be written down explicitly 
in terms of elliptic functions, see for example (\ref{lamkap}), (\ref{abjmloop}) for the case in which the two gauge groups 
have the same rank. It interpolates smoothly between weak and strong coupling, and at strong coupling it agrees precisely with the 
AdS theory prediction obtained in \cite{dp}. 

The 1/6 BPS Wilson loop operator that we study in the planar limit involves only one of the gauge connections in the $U(N_1) \times U(N_2)$ quiver, and 
it is not the most natural operator from the point of view of topological string theory. It has been recently shown in \cite{dt} that it is possible to construct a 1/2 BPS Wilson loop operator in the ABJM theory which localizes precisely to the natural Wilson loop operator for 
Chern--Simons theory on $L(2,1)$ (extended to a supergroup). The vev of such an operator can be calculated 
to all orders in $1/N$, as an exact function of the 't Hooft parameters, by combining the solution of \cite{akmv,hy} with the matrix model-inspired techniques of \cite{mmopen,bkmp}. In the topological string dual, the vevs of these 1/2 BPS operators correspond to open topological string amplitudes. 

Our exact formulae for the vevs of Wilson loops in ABJM theory have the flavour of mirror symmetry. 
There are two sets of coordinates for the parameter space: the 
``bare" coordinates and the flat coordinates. The flat coordinates are identified with the 't Hooft parameters. They are computed by 
period integrals and they can be related to the 
bare coordinates through the mirror map. The vev of the Wilson loop is naturally expressed in terms of bare coordinates, and one has to invert the mirror 
map in order to re-express it in terms of 't Hooft parameters. In retrospect, one could say that the result of \cite{esz,dg,pestun} for the 1/4 BPS Wilson loop in 
$\CN=4$ super Yang--Mills theory is comparatively easier since in their case the relevant matrix model is the Gaussian one, with a simpler moduli space, and where the 't Hooft parameter is simply equal to the bare coordinate. 

The organization of this paper is as follows. In section 2 we review very briefly the matrix model obtained in \cite{kapustin}. In section 3 we present the solution of the lens space matrix model building on \cite{akmv,hy}. In section 4 we explain the relation between the two matrix models, we derive 
the exact results for the 1/6 BPS Wilson loop in the planar limit, and we study its behavior both at strong and at weak coupling. 
Finally, some conclusions are presented in section 5. 

\sectiono{Matrix model for the ABJM Wilson loop}

The ABJM theory is a quiver Chern--Simons--matter theory in three dimensions with gauge group $U(N)_k \times U(N)_{-k}$ and $\CN=6$ supersymmetry. 
The Chern--Simons actions have couplings $k$ and $-k$, respectively, and the theory contains four bosonic fields $C_I$, $I=1, \cdots, 4$, 
in the bifundamental representation of the 
gauge group. One can consider an extension \cite{abj} 
with a more general gauge group $U(N_1)_k \times U(N_2)_{-k}$. A family of Wilson loops in this theory has been constructed in \cite{dp,cw,rey}, with the structure
\be
\label{wloop}
W_R ={1\over d_R(N_1)} \tr_R \, \CP \, \exp \int \left( \ri A_\mu  \dot x^\mu + {2\pi \over k} |\dot x| M^I_J C_I \bar C^J \right) \rd s
\ee
where $A_\mu$ is the gauge connection in the $U(N_1)_k$ gauge group, $d_R(N_1)$ is the dimension of the represenation $R$ of $U(N_1)$, $x^\mu(s)$ 
is the parametrization of the loop, and 
$M_I^J$ is a matrix determined by supersymmetry. It can be chosen so that, if the geometry of the loop is a line or a circle, four real supercharges are preserved. 
Therefore, we will call (\ref{wloop}) the 1/6 BPS Wilson loop. A similar construction exists for a loop based on the other gauge group, 
\be
\label{wloopnar}
\widehat W_R ={1\over d_R(N_2)} \tr_R \, \CP \, \exp \int \left( \ri \widehat A_\mu \dot x^\mu + {2\pi \over k} |\dot x| M_J^I\bar  C_I C^J \right) \rd s 
\ee
where $A_\mu$ is the $U(N_2)_{-k}$ gauge connection. The planar limit of the vev of (\ref{wloop}) was computed in \cite{dp,cw,rey}, for 
$N_1=N_2=N$, in the fundamental representation $R=\tableau{1}$, and in the weak coupling regime $\lambda\ll 1$, where 
\be
\lambda={N\over k}
\ee
is the 't Hooft parameter. The result is
\be
\label{pertw}
\langle W_{\tableau{1}} \rangle =1+  {5 \pi^2 \over 6}\lambda^2 +\CO\left( \lambda^3\right). 
\ee
On the other hand, in the strong coupling regime $\lambda\gg 1$, the Wilson loop vev can be calculated by using the large $N$ string dual, i.e. type IIA 
theory on ${\rm AdS}_4\times \IP^3$ \cite{dp,cw,rey}. This gives the 
prediction\footnote{To be precise, the dual calculation is made by considering a fundamental string in AdS$_4$, 
which is a 1/2 BPS object. In \cite{dp,rey} it was  argued that the the strong coupling behavior obtained in this way should apply to the symmetric Wilson loop $W_{\tableau{1}}^{\rm sym}$ defined in (\ref{sumw}) below. It should also apply to the 1/2 BPS Wilson loop constructed in \cite{dt}. However, as we will show in this paper, 
the leading exponential behavior is common to all these Wilson loops.}
\be
\label{strongw}
\langle W_{\tableau{1}} \rangle \sim \re^{\pi {\sqrt{2\lambda}}}. 
\ee
As in the case of the 1/2 BPS Wilson loop in $\CN=4$ Yang--Mills theory, the exact answer for the planar limit of this vev should interpolate between the weak coupling behavior (\ref{pertw}) 
and the strong coupling prediction of the large $N$ string dual, (\ref{strongw}). 

A crucial step in finding such an exact answer was taken in the paper \cite{kapustin}. It was shown there, through a beautiful application of the localization 
techniques used in \cite{pestun}, that the vev of (\ref{wloop}) can be computed 
as a correlation function in a matrix model. This matrix model is defined by the partition function 
\be
\label{kapmm}
Z_{\rm ABJM}(N_1, N_2, g_s)=\int \prod_{i=1}^{N_1}\rd \mu_i \prod_{j=1}^{N_2} \rd \nu_j {\prod_{i<j} \sinh^2 \left( {\mu_i -\mu_j \over 2}\right)  \sinh^2 \left( {\nu_i -\nu_j \over 2}\right) \over 
\prod_{i,j}  \cosh^2 \left( {\mu_i -\nu_j \over 2}\right)} \re^{-{1\over 2g_s}\left(  \sum_i \mu_i^2 -\sum_j \nu_j^2\right)}, 
\ee
where the coupling $g_s$ is related to the Chern--Simons coupling $k$ of the ABJM theory as
\be
g_s={2 \pi \ri \over k}.
\ee
One of the main results of \cite{kapustin} is that 
\be
\label{16WL}
\langle W_R\rangle ={1\over d_R(N_1)} \left\langle \tr_R\left (\re^{\mu_i}\right)  \right\rangle_{\rm ABJM},
\ee
i.e. the vev of the $1/6$ BPS Wilson loop (\ref{wloop}) can be obtained by calculating the vev of the matrix $\re^{\mu_i}$ in the matrix model (\ref{kapmm}). This 
was explicitly checked in \cite{kapustin} by computing the vev in the r.h.s. of (\ref{16WL}) in the matrix model, for the 
fundamental representation. Notice that the Wilson loop 
for the other gauge group, 
\be
\label{other}
\langle \widehat W_R\rangle ={1\over d_R(N_2)} \left\langle \tr_R \left( \re^{\nu_i} \right) \right\rangle_{\rm ABJM}
\ee
is obtained from (\ref{16WL}) simply by exchanging $N_1\leftrightarrow N_2$ and changing the sign of the coupling constant $g\rightarrow -g$. 

The Wilson loop (\ref{wloop}) breaks the symmetry betwen the two gauge groups. Recently, a class of 1/2 BPS Wilson loops has been constructed in the 
ABJM theory \cite{dt} which treats the two gauge groups in a 
more symmetric way. These loops have a natural supergroup structure in which the quiver gauge group $U(N_1) \times U(N_2)$ is promoted 
to $U(N_1|N_2)$, and they can be defined in any super-representation $\CR$. 
In \cite{dt} it has been argued that this 1/2 BPS loop, which we will denote by ${\rm S}W_\CR$, localizes to the matrix model correlator
\be
\label{12wl}
\langle {\rm S} W_\CR \rangle={1\over s_\CR} \left\langle {\rm Str}_\CR \begin{pmatrix}\re^{\mu_i} &0 \\ 0& -\re^{\nu_j} \end{pmatrix} \right\rangle_{\rm ABJM}
\ee
in the ABJM matrix model. Here, 
\be
s_\CR ={\rm Str}_\CR \begin{pmatrix}1 &0 \\ 0& -1 \end{pmatrix}. 
\ee
When $\CR=\tableau{1}$, we have the simple relationship 
\be
\label{sumw}
\langle {\rm S} W_{\tableau{1}}\rangle = {1\over N_1+N_2} \left(N_1 \langle W_{\tableau{1}}\rangle+ N_2  \langle \widehat W_{\tableau{1}}\rangle \right).
\ee
In general, as it is clear from (\ref{12wl}), the vev of the 1/2 BPS Wilson loop can be obtained if we know the vevs of the 1/6 BPS Wilson loop, but we expect 
it to be simpler. 

The work of \cite{kapustin} reduced the computation of vevs of Wilson loops in ABJM theory to the computation of matrix model correlators in the matrix model (\ref{kapmm}). Perturbative calculations are now straightforward. However, in order to obtain exact interpolation functions, we have to resum all double-line diagrams at fixed genus in the matrix model. This is straightforward in the Gaussian matrix model which computes the vev of the 1/2 BPS Wilson loop in $\CN=4$ SYM \cite{esz,dg,pestun}, but it is not for the model (\ref{kapmm}). As we will see in this paper, the most efficient way to solve this matrix model in the 
$1/N$ expansion is to relate it to the lens space matrix model of \cite{mm} and to its large $N$ string dual \cite{akmv}.

 \sectiono{The lens space matrix model and its large $N$ dual}

The lens space matrix model was introduced in \cite{mm,akmv} in order to compute the partition function of Chern--Simons theory on lens spaces of the form $L(p,1)=\IS^3/\IZ_p$. In this paper 
we will be particularly interested in the case $p=2$. In this case, the matrix model has the structure
\be
\label{intdef}
\ba
Z_{L(2,1)}(N_1, N_2, g_s)=\int\prod_{i=1}^{N_1}\rd \mu_i \prod_{j=1}^{N_2} \rd \nu_j & \prod_{i<j}  \sinh^2 \left( {\mu_i -\mu_j \over 2}\right)  \sinh^2 \left( {\nu_i -\nu_j \over 2}\right) \\
\times &\prod_{i,j}  \cosh^2 \left( {\mu_i -\nu_j \over 2}\right)\,  \re^{-{1\over 2g_s}\left(  \sum_i \mu_i^2 +\sum_j \nu_j^2\right)}
\ea
\ee
and it describes the expansion of the Chern--Simons partition function around a generic non-trivial flat connection, corresponding to the 
symmetry breaking pattern 
\be
U(N) \rightarrow U(N_1) \times U(N_2). 
\ee
The model has a large $N$ expansion of the form
\be
\label{largen}
F=\log Z =\sum_{g=0}^\infty F_g(t_1, t_2)g_s^{2g-2}
\ee
where
\be
t_i=g_s N_i
\ee
are the partial 't Hooft parameters.The genus zero free energy has the structure
\be
\label{gzerotwo}
F_0(t_1, t_2)=F_0^{\rm G}(t_1)+F_0^{\rm G}(t_2) +F_{\rm 0}^{\rm p}(t_1, t_2).
 \ee
where $F^{\rm G}_0(t)$ 
is the Gaussian matrix model genus zero amplitude, 
\be
F_0^{\rm G}(t)={1\over 2} t^2 \Bigl( \log \, t -{3\over 4} \Bigr).
\ee
and $F_{\rm 0}^{\rm p}(t_1, t_2)$ is the contribution from fatgraphs of genus zero. The first nontrivial terms are 
 \be
 \ba
 F_0^{\rm p}(t_1, t_2)&={1\over 288} (t_1^4+6 t_1^3 t_2+18 t_1^2 t_2^2+6 t_1 t_2^3+t_2^4)\\
 &-{1\over 345600} (4 t_1^6+45 t_1^5 
t_2+225 t_1^4 t_2^2+1500 t_1^3 t_2^3+225 t_1^2 t_2^4+45 t_1 t_2^5+4 
t_2^6)+\cdots
\ea
\ee
Higher genus free energies can be computed analogously. 

\begin{figure}[!ht]
\leavevmode
\begin{center}
\includegraphics[height=7cm]{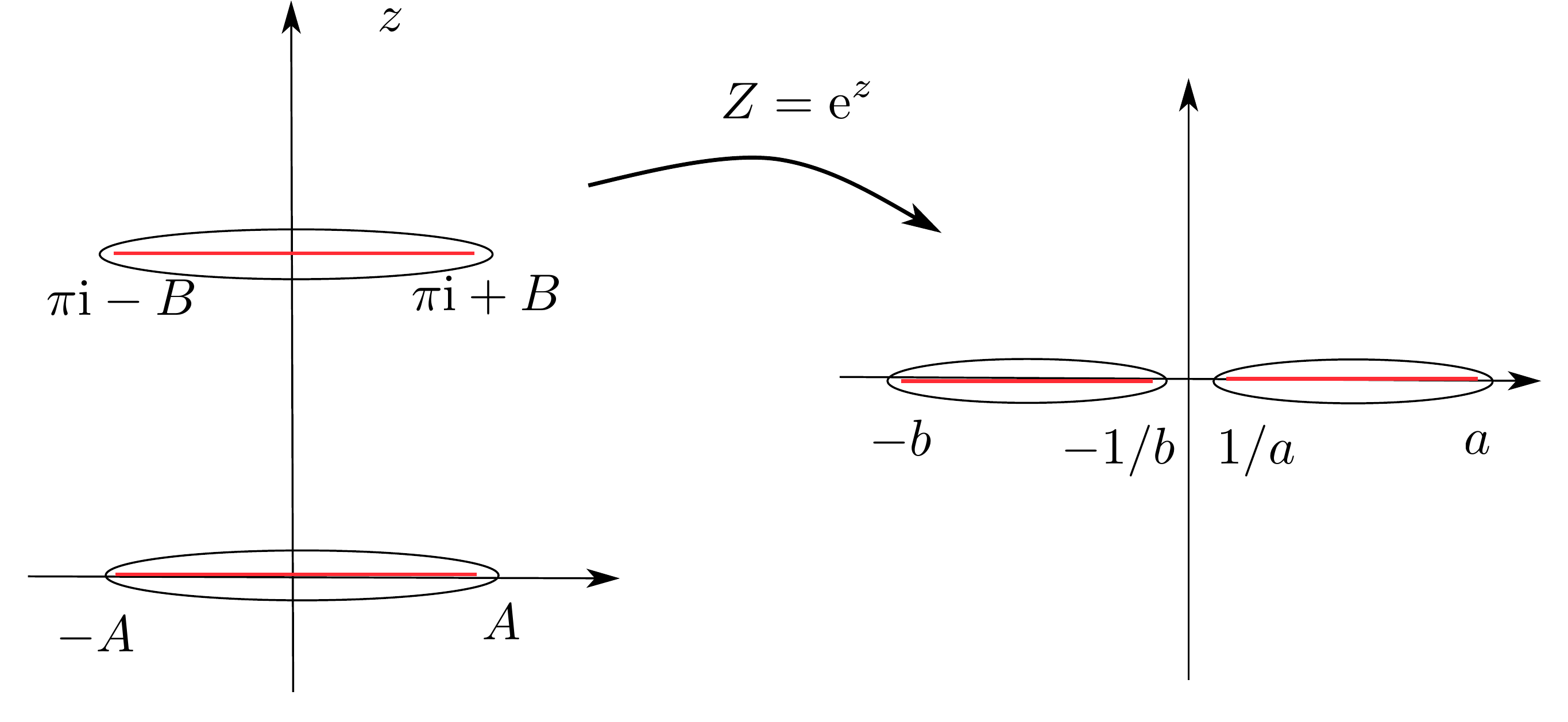}
\end{center}
\caption{The cuts for the CS lens space matrix model in the $z$ plane and in the $Z=\re^z$ plane.}
\label{cuts}
\end{figure}

The matrix model (\ref{intdef}) can be studied from the point of view of the $1/N$ expansion. This was done in detail in \cite{akmv}, where this expansion was 
identified with the genus expansion of topological string theory on a local Calabai--Yau manifold. Many results of \cite{akmv} 
were rederived in the paper \cite{hy,hoy} by using standard matrix model techniques, which we now review. 

At large $N$, 
the two sets of eigenvalues, $\mu_i$, $\nu_j$, condense around two cuts centered around $z=0$, $z=\pi \ri$, respectively. We will write them as
\be
\label{ABcuts}
\CC_1=(-A, A), \qquad \CC_2=(\pi \ri -B, \pi \ri +B),
\ee
in terms of the endpoints $A, B$. It is also useful to use the exponentiated variable 
\be
Z=\re^z, 
\ee
In the $Z$ plane the cuts (\ref{ABcuts}) get mapped to 
\be
(1/a, a), \qquad (-1/b, -b) , 
\ee
which are centered around $Z=1$, $Z=-1$, respectively, and 
\be
\label{Zend}
a=\re^A, \qquad b =\re^B, 
\ee
see \figref{cuts}. We will use the same notation $\CC_{1,2}$ for the cuts in the $Z$ plane. An important quantity introduced in \cite{hy} is the total resolvent of the matrix model, $\omega(z)$. It is defined as
\be
\omega(z) =g_s \left\langle \sum_{i=1}^{N_1} \coth \left( {z-\mu_i \over 2} \right) \right\rangle +g_s \left\langle\sum_{j=1}^{N_2}  \tanh \left( {z-\nu_j \over 2} \right)\right\rangle.
\ee
In terms of the $Z$ variable, it is given by
\be
\omega(z)\rd z =-(t_1 +t_2) {\rd Z  \over Z}  +2  g_s\left\langle \sum_{i=1}^{N_1}{\rd Z \over Z-\re^{\mu_i}}\right\rangle+ 2 g_s \left\langle \sum_{j=1}^{N_2}{\rd Z \over Z+\re^{\nu_j}}\right\rangle\ee
and it has the following expansion as $Z\rightarrow \infty$
\be
\label{Zex}
\omega(z) \rightarrow t_1+ t_2 + {2 g_s  \over Z}\left \langle \sum_{i=1}^{N_1} \re^{\mu_i} -\sum_{j=1}^{N_2} \re^{\nu_j}\right\rangle+\cdots
\ee
From the total resolvent it is possible to obtain the density of eigenvalues at the cuts. In the planar approximation, we have that
\be
\label{planarres}
\omega_0(z) =-(t_1 +t_2) +2  t_1 \int_{\CC_1} \rho_1(\mu) {Z \over Z-\re^{\mu}}\rd \mu + 2  t_2 \int_{\CC_2} \rho_2 (\nu) {Z \over Z+\re^{\nu}}\rd \nu, 
\ee
where $\rho_1(\mu)$, $\rho_2(\nu)$ are the densities of eigenvalues on the cuts $\CC_1$, $\CC_2$, respectively, normalized as
\be
\int_{\CC_1} \rho_1(\mu) \rd \mu=\int_{\CC_2} \rho_2 (\nu) \rd \nu=1.
\ee
The standard discontinuity argument tells us that 
\be
\label{densities}
\ba
\rho_1(X) \rd X &=-{1\over 4 \pi \ri t_1}{\rd X \over X} \left( \omega_0(X+\ri \epsilon) -\omega_0(X-\ri \epsilon)\right), \qquad X\in \CC_1,\\
\rho_2(Y) \rd Y &={1\over 4 \pi \ri t_2 } {\rd Y \over Y}\left( \omega_0(Y+\ri \epsilon) -\omega_0(Y-\ri \epsilon)\right), \qquad Y\in \CC_2.
\ea
\ee

The planar resolvent (\ref{planarres}) was found in explicit form in \cite{hy}. It reads, 
\be
\label{explicitRes}
\omega_0(z) =2 \log \biggl( {\re^{-t/2} \over 2} \Bigl[ {\sqrt{(Z+b)(Z+1/b)}}-{\sqrt{(Z-a)(Z-1/a)}}\Bigr]\biggr),  
\ee
where 
\be
t=t_1+t_2
\ee
is the total 't Hooft parameter. It is useful to introduce the variables 
\be
\alpha=a+{1\over a}, \quad \beta=b+{1\over b}. 
\ee
as well as
\be
\label{zetadef}
\zeta={\alpha-\beta \over 2}, \qquad \xi={\alpha+\beta \over 2}.
\ee
The expansion (\ref{Zex}) implies then 
\be
\xi= 2 \re^{t}. 
\ee
As it is standard in matrix models, the 't Hooft parameters turn out to be period integrals with a nontrivial relation to $a$, $b$:
\be
\label{tperiods}
t_1 ={1\over 4\pi \ri} \oint_{\CC_1} \omega_0 (z)  \rd z , \qquad t_2= {1\over 4\pi \ri} \oint_{\CC_2} \omega_0 (z) \rd z.
\ee
The derivatives of these periods can be calculated in closed form by adapting a trick from \cite{bt}. If we write
\be
\omega_0(z) = \log\left[ {\re^{-t}\over 2} \left( f(Z) -{\sqrt{f(Z)^2-\xi^2 Z^2}}\right) \right]
\ee
with 
\be
f(Z)=Z^2 -\zeta Z +1,
\ee
it follows that 
\be
\label{derper}
{\partial t_{1,2} \over \partial \zeta}=-{1\over 4 \pi \ri} \oint_{\CC_{1,2}} {\rd Z \over {\sqrt{f(Z)^2-\xi^2 Z^2}}} =\pm {{\sqrt{ab}} \over \pi (1+ ab)} K(k), 
\ee
where $K(k)$ is the complete elliptic integral of the first kind, and its modulus is given by 
\be
\label{modulus}
k^2={(a^2-1)(b^2-1) \over (1+a b)^2}. 
\ee

The above relationships determine in principle the planar content of the theory. However, 
this matrix model solution is further clarified by considering its large $N$ dual, 
which was in fact discovered before \cite{akmv}. This dual is given by topological string theory on the anti-canonical bundle
of the Hirzebruch surface $\IF_0=\IP^1 \times \IP^1$. The mirror geometry is encoded in a family of elliptic curves $\Sigma$, which can be written as 
\be
\label{mc}
y={z_1 x^2 + x +1 -{\sqrt{(1+x + z_1 x^2)^2-4 z_2 x^2}}\over 2}. 
\ee
Here, $z_1, z_2$ parametrize the moduli space of complex structures, which is the mirror to the enlarged K\"ahler moduli space of 
local $\IF_0$. This moduli space has a very rich structure discussed in \cite{akmv,hkr}. 

The mirror geometry (\ref{mc}) is nothing but 
the spectral curve 
of the lens space matrix model, and it is closely related to 
the resolvent $\omega_0(Z)$. Indeed, one finds that $\omega_0(Z)\sim \log \, y(x)$ provided we identify the variables as
\be
x=-Z z_1^{-1/2}, 
\ee
and 
\be
\zeta={1\over {\sqrt{ z_1}}}, \qquad \xi= 2{\sqrt{z_2\over z_1}}.
\ee

In order to make further contact with the matrix model, one has to look at the moduli space of (\ref{mc}) near the {\it orbifold} point discovered in \cite{akmv}. This is defined 
as the point $x_1=x_2=0$ in terms of the variables:
\be \label{cov} 
 x_1=1-{z_1\over z_2}, \qquad x_2={1\over \sqrt{z_2}\left(1-{z_1\over z_2}\right)}.
  \ee
Using mirror symmetry, we can calculate the periods of $\omega_0(z)$ along the cycles of the spectral curve as solutions to a Picard--Fuchs equation. In terms of 
the coordinates $x_{1,2}$, the Picard--Fuchs system is given by the two operators
\be
\ba
 \mathcal{L}_1&=\frac{1}{4} \left(8-8 x_1+x_1^2\right) x_2 \d_{x_2}+\frac{1}{4} \left(-4+\left(-2+x_1\right){}^2 x_2^2\right) \d^2_{x_2}\\& 
 +\left(-1+x_1\right) x_1^2 \d_{x_1}-x_1 \left(2-3 x_1+x_1^2\right) x_2 \d_{x_1}\d_{x_2}+\left(-1+x_1\right){}^2 x_1^2 \d^2_{x_1},\\
 \mathcal{L}_2&=\left(2-x_1\right) x_2 \d_{x_2}+\left(-1+\left(1-x_1\right) x_2^2\right) \d^2_{x_2}-x_1^2 \d_{x_1}\\ 
 &+2 \left(-1+x_1\right) x_1 x_2 \d_{x_1}\d_{x_2}+\left(1-x_1\right) x_1^2 \d_{x_1}^2.
 \ea
 \ee
A basis of periods near the orbifold point was found in \cite{akmv}. It reads, 
\be
\label{speriods}
\ba
\sigma_1&=-\log(1-x_1)=\sum_{m} c_{m,0} x_1^m,\\
\sigma_2&=\sum_{m,n} c_{m,n} x_1^m x_2^n,\\
% (2 \pi i)^2\,
\CF_{\sigma_2}&=\sigma_2 \log(x_1)+\sum_{m,n} d_{m,n} x_1^m x_2^n \ ,
\ea
\ee
where the coefficients $c_{m,n}$ and $d_{m,n}$ are determined by the following recursions relations
\be
\label{cdrecursion}
\ba
c_{m,n}=&c_{m-1,n}{(n+2-2m)^2\over 4 (m-n) (m-1)},\cr
c_{m,n}=&{1\over n(n-1)} (c_{m,n-2} (n-m-1)(n-m-2)- c_{m-1,n-2} (n-m-1)^2),\\
d_{m,n}=&{d_{m-1,n}(n+2-2m)^2 + 4 (n+1-2m) c_{m,n}+ 4 (2m-n-2) c_{m-1,n}\over 4 (m-n) (m-1)},\cr
d_{m,n}=&{1\over n(n-1)} (d_{m,n-2} (n-m-1)(n-m-2)- d_{m-1,n-2} (n-m-1)^2 \cr &+(2n-2-2m)c_{m-1,n-2}+(2m+3-2n)c_{m,n-2}).
\ea
\ee
The 't Hooft parameters of the matrix model are related to the periods above as
\be
\ba
 t_1&={1\over 4} (\sigma_1+\sigma_2), \quad
  t_2&={1\over 4} (\sigma_1-\sigma_2).
  \ea
  \ee
The remaining period in (\ref{speriods}) might be used to compute the genus zero free energy of the matrix model. Notice that $x_{1,2}$ (or equivalently 
$\zeta, \xi$ as defined in (\ref{zetadef})) are ``bare" coordinates, while $\sigma_{1,2}$ (and therefore $t_{1,2}$) are flat coordinates, annihilated by the 
Picard--Fuchs operators. 

It is now a matter of (computer) routine to calculate the different quantities, like the endpoints of the cuts, as an expansion in the 't Hooft parameters. One obtains, for example,  
\be
a =1+ 2 {\sqrt{t_1}} +2 t_1 +{1\over 2}  {\sqrt{t_1}} (3 t_1 +t_2) + t_1(t_1+t_2) +\cdots \ee
The expansion for $b$ is obtained from this one simply by exchanging $t_1 \leftrightarrow t_2$. 

\sectiono{Exact results for the ABJM model}
 
 \subsection{General results}
 
We will now use our knowledge of the solution of the lens space matrix model to solve the ABJM model, at least at the planar level. It is clear that the matrix model (\ref{kapmm}) 
is closely related to (\ref{intdef}), but there are some obvious differences: in (\ref{kapmm}) the interaction between the $\mu$ and the $\nu$ eigenvalues is in the denominator, and 
the Gaussian action for the $\nu$s has the opposite sign. These ingredients are precisely the ones needed to make (\ref{kapmm}) a {\it supergroup} extension of (\ref{intdef}). We will now quickly review some results on supermatrix models, following \cite{agm,yost,dv}. A Hermitian supermatrix has the form
\be
\Phi=\begin{pmatrix} A & \Psi \\ \Psi^{\dagger} & C\end{pmatrix}
\ee
where $A$ ($C$) are $N_1\times N_1$ ($N_2 \times N_2$) Hermitian, Grassmann even matrices, and $\Psi$ is a complex, Grassmann odd matrix. The supermatrix model is defined by the partition function 
\be
Z_{\rm s} (N_1|N_2)=\int {\cal D} \Phi \, \re^{-{1\over g_s} {\rm Str} V(\Phi)}
\ee
where we consider a polynomial potential $V(\Phi)$, and ${\rm Str}$ is the supertrace 
\be
{\rm Str}\, \Phi=\tr \, A -\tr \, C.
\ee
 There are two types of supermatrix models with supergroup symmetry 
$U(N_1|N_2)$: the ordinary supermatrix model, and the physical supermatrix model \cite{yost}. The ordinary supermatrix model is obtained by 
requiring $A$, $C$ to be real Hermitian matrices, while the physical model is obtained by requiring that, after diagonalizing $\Phi$ by a superunitary 
transformation, the resulting eigenvalues are real. Here we will be interested in the physical supermatrix model. 
Its partition function reads, in terms of eigenvalues \cite{yost,dv}
\be
\label{supergroup}
Z_{\rm s} (N_1|N_2)=\int \prod_{i=1}^{N_1}\rd \mu_i \prod_{j=1}^{N_2} \rd \nu_j {\prod_{i<j} \left( \mu_i -\mu_j \right)^2 \left( \nu_i -\nu_j \right)^2 \over 
\prod_{i,j}  \left( \mu_i -\nu_j \right)^2} \re^{-{1\over g_s}\left(  \sum_i V(\mu_i) -\sum_j V(\nu_j)\right)}.
\ee
When the two groups of eigenvalues $\mu_i$, $\nu_j$ are expanded around two different critical points, the partition function (\ref{supergroup}) 
is well-defined as an asymptotic 
expansion in $g_s$. It is easy to show that (\ref{supergroup}) is related to the partition 
function of the corresponding bosonic, two-cut matrix model 
\be
Z_{\rm b} (N_1, N_2)=\int \prod_{i=1}^{N_1}\rd \mu_i \prod_{j=1}^{N_2} \rd \nu_j \prod_{i<j} \left( \mu_i -\mu_j \right)^2 \left( \nu_i -\nu_j \right)^2 
\prod_{i,j}  \left( \mu_i -\nu_j \right)^2 \re^{-{1\over g_s}\left(  \sum_i V(\mu_i) +\sum_j V(\nu_j)\right)} 
\ee
after changing $N_2\rightarrow -N_2$: 
\be
\label{bossuper}
Z_{\rm s}(N_1|N_2) =Z_{\rm b}(N_1, -N_2). 
\ee
Such a flip of sign is trivially performed if one knows the exact solution of the model in the $1/N$ expansion. The relation (\ref{bossuper}) can be 
proved diagramatically by introducing Faddeev--Popov ghosts as in \cite{dv,dgkv}. 

We now see that the relationship between the ABJM matrix model and the lens space matrix model is identical to the one we have between supergroup matrix models and multi-cut 
bosonic matrix models, with the only difference that the interaction between the eigenvalues has been promoted to the $\sinh$ interaction typical of Chern--Simons matrix models. 
Indeed, the lens space matrix model is a two-cut matrix model where the $\mu$, $\nu$ eigenvalues are expanded around two different saddle points, $z=0$ and $z=\pi \ri$. The 
ABJM matrix model is just its supergroup version. We then conclude that
\be
\label{changesignZ}
Z_{\rm ABJM}(N_1, N_2, g) =Z_{L(2,1)}(N_1, -N_2, g). 
\ee
Notice that the change $N_2 \rightarrow -N_2$ is equivalent to setting
\be
\label{conABJ}
t_1=2\pi \ri \lambda_1, \qquad t_2=-2\pi \ri \lambda_2,
\ee
where
\be
\label{thooftABJ}
\lambda_i={N_i\over k}, \qquad i=1,2, 
\ee
are the 't Hooft parameters of the ABJM model. 

The appearance of a hidden supergroup structure in the matrix model of \cite{kapustin} is not surprising, since $\CN=4$ Chern--Simons--matter theories 
are classified by supergroups \cite{gw}. In fact, the ABJM theory can be constructed as an $\CN=4$ theory with supergroup $U(N_1|N_2)$ and containing 
both hypermultiplets and twisted hypermultiplets \cite{hosomichi}. This hidden supergroup structure in the ABJM theory is explicitly used 
in the construction of half-BPS Wilson loops in \cite{dt}. 

Let us now discuss Wilson loops. The most natural correlator in the standard lens space matrix model is 
\be
\label{correl}
 \left \langle\tr_\CR\begin{pmatrix} \re^{\mu_i} & 0\\ 0& -\re^{\nu_j} \end{pmatrix} \right\rangle,
\ee
where $\CR$ is a representation of $U(N_1+N_2)$. In Chern--Simons gauge theory on $L(2,1)$, this computes the vev of a trivial knot, expanded around a 
generic, fixed flat connection. In the topological string large $N$ dual, it corresponds to an {\it open} string amplitude for a toric D-brane (see for example \cite{bkmp} for 
more details). These vevs can be computed for any $\CR$ and to all orders in the $1/N$ expansion \cite{mmopen,bkmp}. 

In order to consider its supergroup extension, notice that a representation of $U(N_1+N_2)$ induces a super-representation of $U(N_1|N_2)$, defined by the 
same Young tableau $\CR$ (see 
for example \cite{bars}). Therefore,  the supergroup generalization of (\ref{correl}) is simply
\be
\label{supercor}
 \left \langle{\rm Str}_\CR\begin{pmatrix} \re^{\mu_i} & 0\\ 0& -\re^{\nu_j} \end{pmatrix} \right\rangle_{(N_1|N_2)}. 
\ee
This can be also written as \cite{bars}
\be
\ba
{\rm Str}_\CR\begin{pmatrix} \re^{\mu_i} & 0\\ 0& -\re^{\nu_j} \end{pmatrix}& =\sum_{\vec k} {\chi_\CR(\vec k) \over z_{\vec k}}  \prod_\ell 
\left( {\rm Str}\begin{pmatrix} \re^{\mu_i} & 0\\ 0& -\re^{\nu_j} \end{pmatrix}^\ell \right)^{k_\ell}\\
&=\sum_{\vec k} {\chi_\CR(\vec k) \over z_{\vec k}} 
\prod_\ell \left( \tr  \left( \re^{\ell \mu_i}\right) -(-1)^\ell \tr \left( \re^{\ell \nu_j} \right) 
\right)^{k_\ell}.
\ea
\ee
In this equation, which is the supergroup generalization of Frobenius formula, $\vec k=(k_\ell)$ is a vector of non-negative, integer entries, which can be regarded as a conjugacy class of the symmetric group, $\chi_\CR(\vec k)$ is the character of this conjugacy class in the representation $\CR$, and 
\be
z_{\vec k}=\prod_\ell \ell^{k_\ell} k_\ell!
\ee
We then have 
\be
\label{vevrels}
\left \langle{\rm Str}_\CR\begin{pmatrix} \re^{\mu_i} & 0\\ 0& -\re^{\nu_j} \end{pmatrix} \right\rangle_{(N_1|N_2)}=\left \langle\tr_\CR\begin{pmatrix} \re^{\mu_i} & 0\\ 0& -\re^{\nu_j} \end{pmatrix} \right\rangle(N_1, -N_2, g)
\ee
which extends (\ref{changesignZ}) to correlation functions. In view of (\ref{12wl}), we conclude that the vevs of the 1/2 BPS Wilson loops constructed 
in \cite{dt} can be computed to all orders in the $1/N$ expansion by calculating the correlator (\ref{correl}) in the lens space matrix model and changing $N_2\rightarrow -N_2$
\be
\label{vevsym}
\left\langle {\rm S} W_\CR \right \rangle={1\over s_R} \left \langle\tr_\CR\begin{pmatrix} \re^{\mu_i} & 0\\ 0& -\re^{\nu_j} \end{pmatrix} \right\rangle(N_1, -N_2, g).
\ee

Let us give an example of how to calculate the vev (\ref{vevsym}) when $\CR=\tableau{1}$. In the planar limit we have, for the lens space matrix model correlator (\ref{correl}), the exact answer
\be
\label{12exact}
 {1\over N_1 -N_2} \left \langle\tr_{\tableau{1}}\begin{pmatrix} \re^{\mu_i} & 0\\ 0& -\re^{\nu_j} \end{pmatrix} \right\rangle ={1\over t_1 -t_2}{\zeta\over 2},
\ee
where $\zeta$ is defined in (\ref{zetadef}). After setting (\ref{conABJ}) we obtain the weak coupling expansion
\be
\langle {\rm S} W_{\tableau{1}} \rangle=1+\ri \pi  (\lambda_1-\lambda_2)-\frac{1}{3} \pi ^2 \left(2 \lambda_1^2-5\lambda_2 \lambda_1+2 \lambda_2^2\right)-\frac{1}{3}\ri \pi ^3 \left(\lambda_1^3-4 \lambda_2 \lambda_1^2
+4 \lambda_2^2\lambda_1-\lambda_2^3\right) +\cdots
\ee
This is of course (up to normalizations) the first term in equation (6.54) of \cite{bkmp}. 
Higher genus corrections can be extracted from the higher genus resolvents $\omega_g(z)$. These in turn can be computed by using standard matrix model techniques applied to the spectral curve (\ref{mc}), as in \cite{mmopen,bkmp}. 

\subsection{The 1/6 BPS Wilson loop} 

If the densities of eigenvalues  $\rho_1(\mu)$ and $\rho_2(\nu)$ given in (\ref{densities}) are known, it is possible to calculate the exact planar limit of 
the correlator
\be
 {1\over N} \langle \tr\, \re^{\mu_i} \rangle= \int_{\CC_1} \rho_1(\mu) \re^{\mu} \rd \mu =\int_{\CC_1} \rho_1(X) X \rd X,  
\ee
as well as of multiple-winding correlators
\be
 {1\over N} \langle \tr\, \re^{n\mu_i} \rangle = \int_{\CC_1} \rho_1(\mu) \re^{n\mu} \rd \mu. 
\ee
We then conclude, in view of (\ref{16WL}), that the planar limit of the vev of a 1/6 BPS Wilson loop is given by
\be
\label{16Wilson}
\langle W_{\tableau{1}} \rangle=\int_{\CC_1} \rho_1(\mu) \re^{\mu} \rd \mu 
\ee
after changing variables as in (\ref{conABJ}). 

The densities $\rho_1(\mu)$ and $\rho_2(\nu)$
can be explicitly calculated from (\ref{densities}) and (\ref{explicitRes}). We find, 
\be
\ba
\rho_1(X)\rd X &={1\over  \pi t_1} \tan^{-1}\left[ {\sqrt{ \alpha X-1-X^2 \over \beta X +1 + X^2}} \right] {\rd X \over X},\\
\rho_2(Y)\rd Y &=-{1\over  \pi t_2} \tan^{-1}\left[ {\sqrt{\beta Y +1 + Y^2\over  \alpha Y-1-Y^2 }} \right]{\rd Y \over Y}.
\ea
\ee
In terms of the variable $x=\log X$ we have
\be
\label{exdensity}
\rho_1(x) ={1\over  \pi t_1} \tan^{-1}\left[ {\sqrt{ \alpha -2 \cosh x \over \beta +2 \cosh x }} \right],
\ee
and a similar expression for $\rho_2(y)$. Notice that, if $t_2=0$, one has $\beta=2$, $\alpha=4\re^t -2$, and $\rho_1(x)$ becomes the 
density of eigenvalues for the matrix model of Chern--Simons theory on $\IS^3$ \cite{leshouches}
\be
\rho_1(x)={1\over \pi t} \tan^{-1} \left[ { {\sqrt{\re^t -\cosh^2 \left({x\over 2}\right)}} \over  \cosh \left({x\over 2}\right) }\right].
\ee

\begin{remark} We can write the density of eigenvalues (\ref{exdensity}) as 
\be
\rho_1(x) =f(x) {\sqrt{A^2-x^2}}, 
\ee
where
\be
f(x) ={1\over  \pi t_1\sqrt{A^2-x^2}} \tan^{-1}\left[ {\sqrt{ \alpha -2 \cosh x \over \beta +2 \cosh x }} \right] =\sum_{k=0}^{\infty} \beta_k x^{2k}.
\ee
We find for example
\be
\beta_0={1\over \pi t_1 A}\tan^{-1}\left[ {\sqrt{ \alpha -2 \over \beta +2 }} \right].
\ee
It is easy to check that $\rho_1(x)$ agrees, in the special case $t_1=-t_2=t$, with the perturbative expansion obtained in \cite{suyama} up to order $10$ in $t$. Our expression 
(\ref{exdensity}) gives then the {\it full} resummation of the expression obtained in \cite{suyama}, and extends it to any $t_1, t_2$. 
\end{remark}

The integral (\ref{16Wilson}) is then given by
\be
 \langle W_{\tableau{1}} \rangle={1\over \pi t_1} I_1, \quad  I_1=\int_{1/a}^a  \tan^{-1}\left[ {\sqrt{ \alpha X -1-X^2  \over \beta X+1+X^2 }}\right] \rd X. 
\ee
This integral is not easy to calculate in closed form, but its derivatives w.r.t. $\zeta$ and $\xi$ can be expressed in terms of elliptic integrals. 
We find
\be
\ba
 \frac{\partial I_1}{\partial \zeta}&=
 \frac{1}{2}\int\limits_{1/a}^a\frac{X \rd X}{\sqrt{({\alpha X-1-X^2})({\beta X+1+X^2})}}=
- \frac{1}{\sqrt{ab}\,(1+ab)}\left(a\,K(k)-(a+b)\,\Pi(n|k)\right),\\
 \frac{\partial I_1}{\partial \xi}&=
 \frac{1}{2}\int\limits_{1/a}^a\frac{\left(({\beta X+1+X^2})-({\alpha X-1-X^2})\right)\, \rd X}{\sqrt{({\alpha X-1-X^2})({\beta X+1+X^2})}}=
 \frac{\sqrt{ab}}{a+b}\,E(k),
 \ea
  \label{I_1_zeta}
\end{equation}
where $\Pi(n|k)$ is the complete elliptic integral of the third kind, $K(k), E(k)$ are elliptic integrals of the first and second kind, respectively, the modulus is given by (\ref{modulus}), 
and finally 
\begin{equation}
\label{nab}
 n={b \over a} \frac{a^2-1}{1+a b}\;.
\end{equation} 
This determines the planar limit of the 1/6 BPS Wilson loop exactly. As we mentioned in the introduction, the Wilson loop is naturally expressed in terms of the 
``bare" coordinates $\zeta, \xi$, and we have to use the mirror map (\ref{speriods}) in order to write it in terms of the 't Hooft parameters. 

As an application of these formulae, we will present the first few terms of the weak coupling expansion in $t_1, t_2$. To do this, we simply calculate
\be
{\partial I_1 \over \partial t_i} = \frac{\partial I_1}{\partial \zeta}{\partial \zeta \over \partial t_i} +  2 \frac{\partial I_1}{\partial \xi} \re^t,
\ee
we use the expressions (\ref{I_1_zeta}), and we integrate w.r.t. $t_1$, $t_2$. This produces the expansion
\be
 \langle W_{\tableau{1}} \rangle=1+ {t_1\over 2} + {1\over 12}(2 t_1^2 + 3 t_1 t_2) + {1\over 48}(2 t_1^3 + 6 t_1^2 t_2 + 4 t_1 t_2^2) +{1\over 960}  (8 t_1^4 + 35 t_1^3 t_2 + 30 t_1^2 t_2^2 + 10 t_1 t_2^3)+\cdots
\ee
The result for the Wilson loop in the ABJ theory is obtained by simply changing variables to (\ref{conABJ}). 
The result agrees at the first few orders with a perturbative matrix model calculation of \cite{dt}. Finally, the vev of the other Wilson loop (\ref{other}) is obtained by 
exchanging $t_1 \leftrightarrow t_2$, or, in the ABJ theory, by exchanging $\lambda_1 \leftrightarrow \lambda_2$ and complex conjugating the result. 

Interestingly, the strong coupling limit of the above expressions depends on the direction in which one goes to infinity. We will discuss one such direction in detail in the 
next subsection, where we consider the restriction to the original ABJM model $N_1=N_2$. 

\subsection{The ABJM slice}

In the original ABJM theory with $N_1=N_2=N$ (the case $N_1\not= N_2$ was considered in \cite{abj}) we should look at the slice 
\be
\label{abjmslice}
t_1=-t_2=2\pi \ri \lambda, \quad \lambda={N\over k}
\ee
in the moduli space of the dual topological 
string. From the point of view of the periods $\sigma_1$, $\sigma_2$ in (\ref{speriods}) this means that we should set 
\be
\sigma_1=0, 
\ee
therefore $x_1=0$. In order to have a nontrivial $\sigma_2$, we must consider the double-scaling limit
\be
\label{dsl}
x_1 \rightarrow 0,  \qquad  x_1x_2 =\zeta \quad {\text{fixed}}.
\ee
The one-dimensional subspace (\ref{abjmslice}) corresponds, in terms of the variables $\zeta, \xi$, to $\xi=2$. As in \cite{beyondgenus}, we can find simplified expressions for the 
periods in this subspace. It is easy to see from the structure of $\sigma_2$ that, in the limit (\ref{dsl}), one has
\be
\sigma_2=\sum_{m=0}^{\infty} a_m \zeta^{2m+1}, \quad a_m=c_{2m+1, 2m+1},
\ee
and from the recursion relation (\ref{cdrecursion}) we find
\be
a_m= {2^{-4 m} \Gamma \left( m+{1\over 2}\right)^2 \over \pi (2m+1)  \Gamma \left( m+1\right)^2}.
\ee
We then obtain
\be
{\rd \sigma_2\over \rd \zeta} ={2\over \pi} K\left({\zeta\over 4}\right),
\ee
which is in fact a particular case of (\ref{derper}), as it can be easily seen by using the transformation properties of the 
elliptic integral $K(k)$. The period $t_1$ itself can be written as a generalized hypergeometric function:
\be
\label{tonex}
t_1(\zeta)={\zeta\over 4}
   {~}_3F_2\left(\frac{1}{2},\frac{1}{2},\frac{1}{2};1,\frac{3}{2};\frac{\zeta^2
   }{16}\right). 
   \ee
 In the physical ABJM theory, $t_1$ is purely imaginary. This means that $\zeta$ is purely imaginary as well, so we set 
 \be
 \zeta=\ri \kappa
 \ee
 and we finally obtain 
 \be
 \label{lamkap}
 \lambda(\kappa)={\kappa \over 8 \pi}   {~}_3F_2\left(\frac{1}{2},\frac{1}{2},\frac{1}{2};1,\frac{3}{2};-\frac{\kappa^2
   }{16}\right).
   \ee

   Let us now calculate the planar limit of the vev of the 1/6 BPS Wilson loop in the ABJM slice $N_1=N_2$. Since
   \be
   \alpha=2+\ri \kappa, \qquad \beta =2-\ri \kappa, 
   \ee
the endpoints of the cuts are given by 
 \be
 \ba
  a(\kappa)&=\frac{1}{2} \left(2+ \ri \kappa+\sqrt{\kappa(4 \ri -\kappa)} \right), \\ 
  b(\kappa)&=\frac{1}{2} \left(2-\ri \kappa+\sqrt{-\kappa(4\ri +\kappa)} \right).
  \ea
  \ee
 The planar vev of the Wilson loop is then determined, as a function of the 't Hooft coupling $\lambda$, by the single equation
   \be
   \label{abjmloop}
  {\rd \over \rd \kappa} \left(\lambda(\kappa)\langle W_{\tableau{1}}\rangle\right) =- \frac{1}{2 \pi^2 \sqrt{ab}\,(1+ab)}\left(a\,K(k)-(a+b)\,\Pi(n|k)\right), 
  \ee
  together with the explicit relation between $\lambda$ and $\kappa$ in (\ref{lamkap}) --yet another example of mirror map. 
  
  As a check, we can perform a weak coupling expansion. The weakly coupled region 
  corresponds to 
  \be
  \kappa\ll 1, \qquad \lambda\ll 1,
  \ee
 and in this region the variables are related as
 \be
 \label{mmap}
{\kappa \over 8 \pi} = \lambda +\frac{\pi ^2 \lambda ^3}{3}-\frac{7 \pi ^4 \lambda ^5}{60}+\frac{173 \pi ^6 \lambda ^7}{1260}-\frac{37927 \pi ^8
   \lambda ^9}{181440}+\CO\left(\lambda ^{10}\right),
   \ee
   which is obtained from the inversion of (\ref{lamkap}). By expanding (\ref{abjmloop}) in power series around $\kappa=0$, and using the mirror map (\ref{mmap}), we obtain
  \be
  \ba
& \langle W_{\tableau{1}}\rangle=  \re^{\pi \ri \lambda}\biggl( 1+\frac{5 \pi ^2 \lambda ^2}{6}-\frac{1}{2} \ri \pi ^3 \lambda ^3-\frac{29
   \pi ^4 \lambda ^4}{120}+\frac{1}{12} \ri \pi ^5 \lambda ^5+\frac{151 \pi
   ^6 \lambda ^6}{1008}-\frac{1}{10} \ri \pi ^7 \lambda ^7-\frac{87449 \pi
   ^8 \lambda ^8}{362880}\\ 
   & +\frac{2603 \ri \pi ^9 \lambda
   ^9}{15120}+\frac{3447391 \pi ^{10} \lambda
   ^{10}}{7983360}-\frac{1166161 \, \ri \pi ^{11} \lambda
   ^{11}}{3628800}-\frac{5239372319 \pi ^{12} \lambda
   ^{12}}{6227020800}+\CO\left(\lambda ^{13}\right) \biggr),
   \ea
   \ee
 where we have extracted a framing factor to facilitate comparison with existing results like (\ref{pertw}). The first few terms agree with the calculations in \cite{dp,cw,kapustin,suyama}. 
  
     \FIGURE[ht]{
\leavevmode
\centering
\hspace{3cm}
\includegraphics[height=5cm]{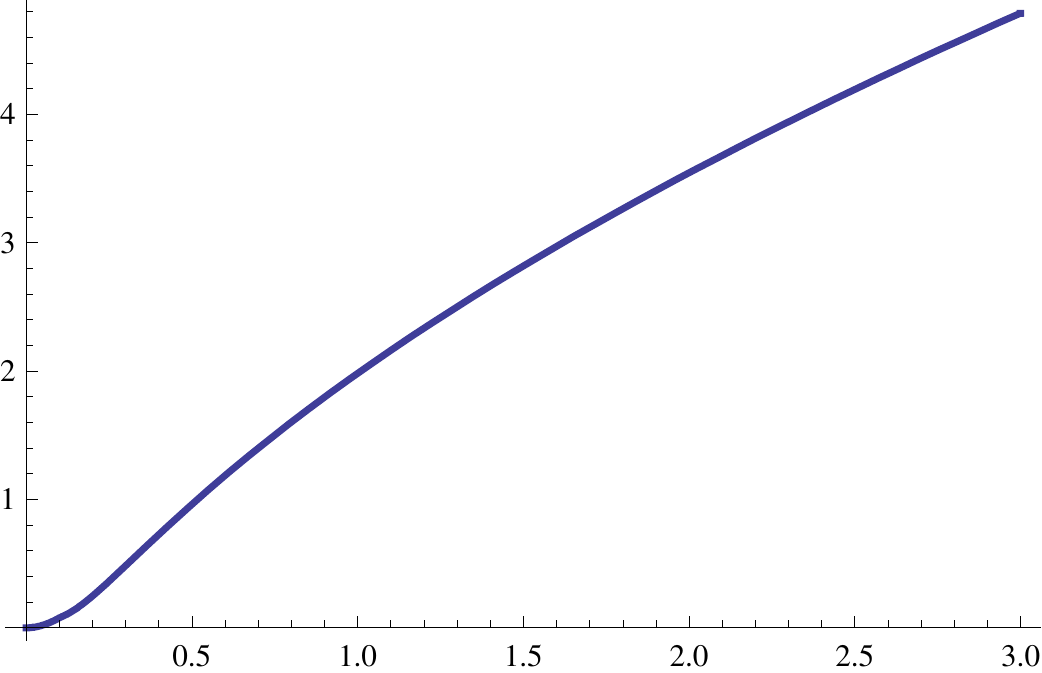}
\hspace{3cm}
\caption{Plot of (\ref{flam}) as a function of $\lambda$, displaying clearly an interpolation between a quadratic behavior near the 
origin and a square-root growth as $\lambda$ becomes large.}
\label{interpol}
}
 Of course, the main advantage of having analytic expressions is that one can perform a weak-strong coupling interpolation easily. 
 The strong coupling region is 
 \be
 \kappa \gg1, \qquad  \lambda\gg 1
 \ee
 and (\ref{lamkap}) leads to the asymptotic expansion
 \be
 \lambda(\kappa) ={\log ^2(\kappa)\over 2 \pi
   ^2}+\frac{1}{24}+\CO\left( {1\over \kappa^2}\right),
   \ee
   which is immediately inverted to 
   \be
   \kappa=\re^{\pi {\sqrt{2\lambda}}}\left(1 +\CO\left({1\over {\sqrt{\lambda}}},  \re^{-2\pi {\sqrt{2\lambda}}} \right) \right).
   \ee
   On the other hand, it is easy to check from (\ref{abjmloop}) that
   \begin{equation}
 \frac{\rd I_1}{\rd\kappa}=-{1\over 2} {\log\kappa}+{\ri \pi \over 4} +\CO\left( {1\over \kappa^2}\right)\;\Rightarrow I_1=-\frac{1}{2}\,\kappa\,\log\kappa+\left( {1\over 2} +{\ri \pi \over 4}\right) \kappa+\CO\left( {1\over \kappa}\right)
\end{equation} 
It follows that
\be
\langle W_{\tableau{1}}\rangle \sim {\ri \over 2 \pi {\sqrt{2\lambda}}} \re^{\pi {\sqrt{2\lambda}}}, \qquad \lambda  \gg 1.
 \ee
 The leading exponential is in perfect agreement with the AdS prediction (\ref{strongw}), and the exact answer interpolates between the weak and the strong coupling behaviors. This is illustrated in \figref{interpol}, which represents the function
 \be
 \label{flam}
 f(\lambda)=\log |\langle W_{\tableau{1}}\rangle|.
\ee
This function interpolates between
 \be
 f(\lambda)=\begin{cases} \displaystyle{{5 \pi^2 \lambda^2 \over 6}},&  \lambda\ll 1,\\
 \\
\displaystyle{ \pi {\sqrt{2\lambda}}}, & \lambda\gg 1.
 \end{cases}
 \ee

Let us now consider the vev of the 1/2 BPS Wilson loop of \cite{dt} in the fundamental representation, 
which is given by (\ref{sumw}). The expression for this vev follows from the specialization of (\ref{12exact}),
\be
\langle {\rm S} W_{\tableau{1}}\rangle ={\kappa \over 8 \pi \lambda(\kappa)}.
\ee
Since, in the ABJM slice, (\ref{16WL}) and (\ref{other}) are related by complex conjugation, this vev is manifestly real. 
Notice also that, when adding the contributions of the 
two gauge groups as in (\ref{sumw}), all terms in the series expansion 
at small $\kappa$ of the r.h.s. of (\ref{abjmloop}) cancel except for the first constant term. This confirms 
that the 1/2 BPS Wilson loop is much simpler than the 1/6 BPS Wilson loop. At strong coupling we have, 
\be
\langle {\rm S} W_{\tableau{1}}\rangle \sim {1 \over 8 \pi \lambda } \re^{\pi {\sqrt{2\lambda}}},  \qquad \lambda  \gg 1,
\ee
which displays the same leading exponential behavior predicted by the large $N$ dual. 

The computations at strong coupling can be easily extended to the case $N_1 \not=N_2$, but they depend on the direction in which we take the limit in the 
space of 't Hooft parameters. 
For $\xi$ fixed and $\zeta$ large (therefore $\lambda_1\sim \lambda_2$), we find the same exponential behavior
\be
 \langle W_{\tableau{1}}\rangle \sim \re^{\pi {\sqrt{2\lambda_1}}} \sim \re^{\pi {\sqrt{\lambda_1 +\lambda_2}}}. 
 \ee

\sectiono{Conclusions}

In this paper we have related the matrix integral of \cite{kapustin}, computing Wilson loop vevs in ABJM theory, to a supergroup extension of 
the lens space matrix model introduced in \cite{mm} and solved in \cite{akmv,hy}. This has made possible to obtain exact expressions for the planar 
vev of 1/6 BPS Wilson loops, and we have verified the strong coupling behavior predicted by the AdS dual \cite{dp,rey,cw}. 

There are various avenues for further research. It is clearly gratifying to see that the techniques of mirror symmetry are relevant to 
the strong-weak interpolation problem in the AdS/CFT correspondence, but it would be nice to have an {\it a priori} understanding of the relationship 
between topological string theory and the ABJM theories, maybe along the lines of \cite{bh}. This would lead to further fruitful interactions between the two topics. 

We should also mention that the strong coupling analysis of Wilson loops 
in the context of the gauge theory is different from what one does in topological string theory. 
There, the interpolation from weak to strong coupling involves a different choice of flat coordinates and a different choice of duality frame (see for example \cite{abk} and specially the analysis of Wilson loop amplitudes in \cite{bkmp}). This means, 
in particular, that in topological string theory, analytic continuation of the amplitudes is not enough. In the gauge theory analysis, in contrast, 
one uses the same flat coordinates (the 't Hooft parameters) and the same duality frame in the full moduli space, and the weak and the strong coupling 
regions are simply related by an analytic continuation.    

One obvious problem is to find expressions for the Wilson loop vevs beyond the planar approximation. 
As we have explained in this paper, this can be trivially done for the 1/2 BPS Wilson loops constructed in \cite{dt}, by simply applying 
the techniques of \cite{mmopen,bkmp}. For the 1/6 BPS Wilson loop the calculation of $1/N$ corrections is more difficult, but in principle it can be done. Finally, it would be interesting to compare the matrix model results with an AdS calculation using 
D-branes, as it was done in \cite{df} for the 1/2 BPS Wilson loop of $\CN=4$ Yang--Mills theory. 

%%%%%%%%%%%%%%%%%%%%%%%%%%%%%%%%%%%%%%%%%%%%%%%%%%%%%%%%%%%%%%%%%
\section*{Acknowledgments}
We would like to thank Anton Kapustin for discussions. We are specially grateful to Nadav Drukker for many conversations which helped us very much, and we want to thank him, as well as Diego Trancanelli, for sharing their unpublished results with us. This work is supported in part by the 
Fonds National Suisse. 

%%%%%%%%%%%%%%%%%%%%%%%%%%%%%%%%%%%%%%%%%%%%%%%%%%%%%%%%%%%%%%%%%

\end{document}